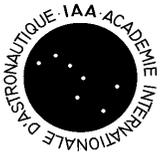
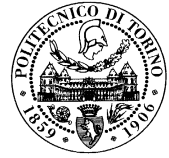

# SCIENCE ENABLED BY A MOON VILLAGE

**Ian A. Crawford**
Department of Earth and Planetary Sciences, Birkbeck College, University of London, UK
i.crawford@bbk.ac.uk

**Abstract**: A human-robotic 'Moon Village' would offer significant scientific opportunities by providing an infrastructure on the lunar surface. An analogy would be the way in which human outposts in Antarctica facilitate research activities across multiple scientific disciplines on that continent. Scientific fields expected to benefit from a Moon Village will include: planetary science, astronomy, astrobiology, life sciences, and fundamental physics. In addition, a Moon Village will help develop the use of lunar resources, which will yield additional longer-term scientific benefits.

## 1. INTRODUCTION

The ESA Director General has suggested that the creation of a human-robotic lunar outpost (or 'Moon village') would be a logical next step in human space exploration [1]. The creation of such an outpost would offer significant scientific opportunities by providing an infrastructure on the lunar surface [2-5]. An analogy would be the way in which human outposts in Antarctica facilitate research activities across multiple scientific disciplines on that continent [6,7]. In the Moon Village case, scientific fields that might be expected to benefit will include: planetary science, astronomy, astrobiology, life sciences, and fundamental physics. In addition, a Moon Village would help assess the economic potential of lunar resources [8,9], while also acting as a 'market' for such resources.

## 2. SCIENCE ENABLED BY A MOON VILLAGE

The extent to which different scientific areas will benefit from a lunar outpost will, in part, depend on its location and on the duration for which it is occupied. Some of the scientific areas discussed below are more dependent on these factors than others. For example, science questions related to the lunar poles would clearly benefit more from a polar Moon Village than an equatorial one. Nevertheless, the kind of transportation and other infrastructure required to establish a Moon Village is likely to facilitate human and robotic operations at locations that may not be local to the Moon Village itself. Moreover, in the fullness of time we might envisage multiple such outposts being established at different locations.

With these caveats in mind, we here address the major areas of science that would benefit from the scientific infrastructure represented by a Moon Village.

## 2.1. Planetary science

As discussed in references [4,5], the lunar geological record still has much to tell us about the earliest history of the Solar System, the origin and evolution of the Earth-Moon system, the geological evolution of rocky planets, and the near-Earth cosmic environment throughout Solar System history. Accessing this record would be greatly enabled by again having humans operating on the lunar surface, and especially if supported by a scientific infrastructure such as that envisaged for the Moon Village. This would be especially true with regard to the range of rock and soil samples that could be collected, analysed (and, if necessary, returned to Earth), and for the implementation of complex exploratory activities such as sub-surface drilling.

Specific planetary science activities that would benefit from such an infrastructure include the following (see references [4,5,10,11], and references cited therein, for additional details):

- *Constraining the bombardment history of the inner solar system*. The lunar cratering rate is used to estimate the ages of cratered surfaces throughout the solar system, but it is not well calibrated (e.g. [12]). Improving this calibration is a key objective of planetary science, and will require sampling impact melt deposits from multiple (ideally hundreds) of impact craters of a wide range of ages. In addition, recovering fragments of the impacting bodies will reveal how and if the population of impactors has changed with time [13]. Accessing these materials will require a significant transportation infrastructure (e.g. pressurised rovers) on the lunar surface, a significant sample return capacity and, ideally, *in situ* analytical instruments.

- *Understanding the impact cratering process*. Impact cratering is a fundamental planetary process, but our understanding of it is limited by lack of access to pristine craters of a wide range of sizes. The lunar surface clearly exhibits the required diversity of craters. Using these craters to improve our knowledge of impact cratering processes will require visiting many craters of a wide range of sizes and ages to conduct sampling, geophysical surveys and, possibly, sub-surface drilling.

- *Determining the structure, composition and evolution of the lunar interior*. The Moon provides an excellent example of a small rocky planet which largely preserves its internal structure from shortly after its formation. As such it can provide insights into the early geological evolution of larger and more complicated planets. To better understand the structure of the lunar interior will require emplacement of geophysical tools (e.g. seismic networks and magnetic surveys), as well as sampling a wide range of lunar crustal and volcanic rocks of diverse ages and compositions. Such studies will also aid in constraining theories of lunar origin.

- *Determining the extent and distribution of lunar volatiles*. There is now convincing evidence that water ice exists within permanently shadowed polar craters, and that hydrated materials also exist at high-latitude (but not permanently shadowed) localities. In addition to their possible value as resources (see below), such deposits will inform our knowledge of the role comets and meteorites have played in delivering volatile substances to the terrestrial planets. Determining the nature and extent of these volatiles will require surface activities (including sampling and near-surface drilling) at polar and high-latitude locations.

- *Understanding regolith processes on airless bodies*. The lunar surface is a natural laboratory for understanding space weathering and regolith processes throughout the solar system. Access to regoliths of different composition, thickness, and latitude will be required to improve our knowledge of these processes.

- *Serendipitous discoveries*. Humans are unique in our ability to recognize new observations or phenomena to be of importance, even if not anticipated in advance. It follows that having humans living and working on the lunar surface for long periods is likely to result in unanticipated discoveries that might not otherwise be made. Although by definition unquantifiable, such discoveries may ultimately prove to be among the most significant to result from a lunar outpost.

## 2.3. Astronomy

The lunar surface is a potentially useful platform for astronomical observations [14-16], and establishing the equipment required for such observations would be facilitated by a human and robotic infrastructure on the Moon. Key aspects include:

- *Low frequency radio astronomy from the lunar farside*. Radio wavelengths longer than about 10m cannot penetrate the Earth's ionosphere, yet much valuable scientific information is expected to lie in this frequency range (e.g. [16]). The lunar farside, which is permanently shielded from the Earth, is probably the best location in the inner solar system for such observations and a human infrastructure on the Moon would assist in the setting up of the necessary equipment (although care will also be required to ensure that human operations do not degrade the natural radio-quietness of the location).

- *Optical and infra-red astronomy*. Although the lunar surface could in principle provide a platform for optical and infra-red telescopes, there is a consensus that free-flying satellites (e.g. at the Earth-Sun L2 point) probably provide better platforms for such activities [17]. It is certainly true that some aspects of the lunar surface environment (e.g. dust and, in most locations, extreme diurnal temperature ranges) are not amenable to astronomical observations at these wavelengths. Nevertheless, access to a scientific infrastructure, able to emplace, maintain and upgrade such instruments (the value of which has been demonstrated by multiple HST servicing missions) might compensate for these disadvantages. Moreover, the stability of the lunar surface might be enabling in the context of building optical/IR interferometric arrays, and permanently shadowed polar craters might enable passive cooling of IR instruments. The pros and cons of the lunar surface for optical/IR astronomy still need to be properly assessed and a lunar outpost would at least facilitate such an assessment.

- *Cosmic-ray observations*. As the lunar surface is not shielded by either an atmosphere or a magnetic field (other than the Sun's heliospheric magnetic field) it is an attractive location from which to study the flux and composition of primary cosmic rays. This can also be done in free space, but the existence of a human and robotic infrastructure on the lunar surface may facilitate the installation of the required instrumentation. A similar argument holds for Gamma- and X-Ray observations.

## 2.3. Astrobiology

Astrobiology is usually defined as the study of the origin, evolution, distribution, and future of life in the universe. As the Moon has (almost certainly) never had any indigenous life of its own, at first sight a scientific lunar infrastructure might not seem especially relevant to astrobiology. However, we can identify the following areas of astrobiological research that would benefit from lunar surface operations (see references [4,18,19], and references cited therein, for additional details):

- *Constraining the impact regime under which life arose and evolved on Earth*. Understanding the extent to which large meteorite impacts may have affected the origin and early evolution of life on Earth is a key aspect of astrobiology. It will be naturally addressed as part of the wider calibration of inner solar system cratering rate discussed above, and will require sampling of ancient impact melt deposits in the age range 4.5-3.8 Gyr ago.

- *Accessing the records of solar and galactic evolution recorded in buried palaeoregolith layers*. As discussed in [18,20,21], buried ancient regoliths ('palaeoregoliths') are expected to contain records of the solar wind and galactic cosmic ray (GCR) fluxes throughout solar system history, both of which will inform our understanding of the past habitability of the Earth. Accessing such deposits will require extensive fieldwork and, possibly, drilling to depths of tens of metres, and is the kind of large-scale exploratory activity that would be greatly facilitated by a human infrastructure on the Moon.

- *Sampling materials from early Earth, Mars and Venus*. As pointed out in references [22,23], the lunar surface may contain fragments of the early Earth (and possibly also of Mars and Venus) that pre-date the oldest existing surfaces on those planets. Finding such material, especially for the early Earth, could help constrain the timing of the origin and early evolution of life on our planet and would be of considerable astrobiological significance. Finding such materials on the Moon would probably require extensive exploratory activities [24] and would be aided by a human infrastructure such as would be provided by a Moon Village.

- *Sampling of Earth's early atmosphere.* There is a possibility that molecules derived from the Earth's atmosphere may be preserved in the lunar regolith [25,26]. If these can be identified in, and extracted from, ancient palaeoregolith layers then they have the potential to provide insights into the evolution of Earth's atmosphere with time. Accessing them will require the same sort of capabilities required for extracting solar wind and GCR records from palaereregolith layers as discussed above.

- *Understanding the astrobiological significance of irradiated ices*. Lunar polar ices exposed to GCRs may be expected to form organic molecules [27]. Sampling of polar ices, as described above, will indicate the extent to which this process occurs and its relevance to abiotic syntheses of organic molecules in planetary and interstellar ices.

- Studying the survival of organic material (including spores and, conceivably, even live micro-organisms) within the remains of space vehicles (including Apollo lunar module ascent stages) that have crashed onto the lunar surface [28]. Sampling such localities, and determining what, if any, viable organic material has survived for decades on the lunar surface will provide key information regarding the survival of life in the space environment relevant to fundamental biology, planetary protection and panspermia.

- *SETI from the Moon.* At the other extreme of the astrobiology spectrum, a scientific infrastructure on the Moon could help constrain the prevalence of technological civilisations in the Galaxy. There are two main possibilities; (i) SETI could make us of radio astronomical facilities established on the Moon [29]; and, more speculatively (ii) the lunar surface might be searched for artefacts of extraterrestrial origin [30,31]. Both would benefit from a human-tended infrastructure on the lunar surface, but the latter (speculative though it is) would probably absolutely require such an infrastructure owing the large surface areas that would need to be searched to place any meaningful limits on the existence of any such artefacts.

- *Preparing for deep-space exploration*. Insofar as astrobiology, broadly considered, includes consideration of the future of life in the universe the prospects for human expansion beyond Earth fall within its remit. Clearly establishing a Moon Village would in itself be a significant step in this direction. Moreover, it would help build knowledge and expertise that will be required if humans are to move further out into the solar system, and especially to the planet Mars. Much of this knowledge and experience will be in the field of the life sciences, to which we now turn.

## 2.4. Life sciences

The lunar surface will provide multiple opportunities for research in the life sciences (e.g. [4,32,33]). Most of these would not be dependent on the physical location of a human-tended outpost, although some would require a prolonged presence. Gronstal et al. [32] have reviewed the kind of laboratory equipment that will be required for such investigations, and it is clear that a human-tended outpost would greatly facilitate this work (and will, of course, be essential for human studies). Specific areas of lunar life science investigations include:

- *Understanding the response of life to the low, but non-zero, gravity*. Although a lot of research has now been performed on the response of living things to microgravity (e.g. [34,35]), no comparable data exists for prolonged exposure to low, but non-zero, gravity. A lunar outpost would permit such studies on life-forms ranging from individual cells to entire organisms (including humans). Fundamental insights into biological processes may be expected from such studies, in addition to knowledge that will aid in the future exploration of space.

- *Understanding the response of life (again ranging from individual cells to entire organisms) to the lunar radiation and dust environment*.

- *Understanding the response of human physiology to the lunar environment*. Specifically human-oriented research will aid in the development of countermeasures to help enable the long-term human habitation of the Moon and other planetary surfaces [35].

- *Agricultural experiments*. If humans are to have a long-term future on the Moon, and on other planetary surfaces, then growing food will at some stage become necessary. A lunar outpost would facilitate the necessary research.

## 2.5. Fundamental physics

Although not a major driver for lunar exploration, several areas of fundamental physics research may benefit from the scientific infrastructure represented by a lunar outpost. These include:

- *Test of General Relativity through improved lunar laser ranging measurements* [36].

- *Tests of quantum entanglement over the Earth-Moon baseline* [37].

- *Searches for exotic particles (e.g. strange quark matter, dark matter, etc) interacting with the lunar surface* [38].

## 3.   LUNAR RESOURCES

There is growing interest in the future use of lunar resources, both to support lunar exploration itself and as a contribution to a developing space economy [9]. A self-sustaining Moon Village will very likely rely on lunar resources (e.g. locally derived water and oxygen [8,9]), but equally the infrastructure provided by an outpost of this kind will facilitate the prospecting for additional resources. Indeed, a Moon Village could help kick-start a space economy by providing a market for commercial space resource companies while at the same time providing infrastructure to support their activities. Although not strictly a scientific benefit of a Moon Village *per se*, there is little doubt that, in the longer term, science will benefit from the development of a space economy built on the use of space resources [39-41].

## 4. CONCLUSION

By analogy with scientific outposts in Antarctica, a human-robotic outpost on the Moon would enable multiple scientific opportunities owing to the lunar surface infrastructure that it would provide. Scientific fields that would be expected to benefit from such an infrastructure include: planetary science, astronomy, astrobiology, the life sciences, and fundamental physics. In addition, a Moon Village would help initiate the utilization of lunar resources, and perhaps help kick-start a space economy. Although beyond the scope of this paper, a Moon Village would also offer important societal and cultural benefits to humanity, especially if, as envisaged, it is developed as a truly global endeavor [42,43]. Moreover, although not explicitly identified as an objective in the current ISECG Global Exploration Roadmap [44], an international Moon Village would clearly be consistent with this roadmap, and more especially with the over-arching aims of the Global Exploration Strategy [45]. The scientific community, and indeed the world community more generally, has the greatest possible interest in seeing this initiative succeed.